\documentstyle[12pt,amsmath]{article}
\numberwithin{equation}{section}

\newcommand{\grgl}{\:\hbox to -0.2pt{\lower2.5pt\hbox{$\sim$}\hss}
           {\raise3pt\hbox{$>$}}\:}
\newcommand{\klgl}{\:\hbox to -0.2pt{\lower2.5pt\hbox{$\sim$}\hss}
           {\raise3pt\hbox{$<$}}\:}
\def\KK{{\rm I\kern -.2em  K}}
\def\NN{{\rm I\kern -.16em N}}
\def\RR{{\rm I\kern -.2em  R}}
\def\ZZ{Z \kern -.43em Z}
\def\QQ{{\rm \kern .25em
             \vrule height1.4ex depth-.12ex width.06em\kern-.31em Q}}
\def\CC{{\rm \kern .25em
             \vrule height1.4ex depth-.12ex width.06em\kern-.31em C}}
\def\ZZZ{Z\kern -0.31em Z}

\begin{document}
\begin{center}
{\bf\LARGE Quantum correlations}\\
\bigskip
{\bf\LARGE in classical statistics}\\
\vspace{1cm}
C. Wetterich\footnote{e-mail: C.Wetterich@thphys.uni-heidelberg.de}\\
\bigskip
Institut  f\"ur Theoretische Physik\\
Universit\"at Heidelberg\\
Philosophenweg 16, D-69120 Heidelberg\\
\vspace{1cm}
\today
\end{center}

\begin{abstract}
Quantum correlations can be naturally formulated in a classical statistical
system of infinitely many degrees of freedom. This realizes the underlying
non-commutative structure in a classical statistical setting.
We argue that the quantum correlations offer a more robust description with respect
to the precise definition of observables.
\end{abstract}

\newpage

\section{Quantum structures in classical \\statistics}
Classical observables commute, quantum mechanical operators do not - this basic
difference reflects itself in a different behavior of classical correlation functions
and quantum correlations. We will argue here that these different properties are connected
to the formulation of the concept of correlation functions rather than to the ``classical''
or ``quantum'' character of the system itself. Quantum correlations can be formulated in
classical statistics just as well as classical correlation functions may be defined in a
quantum system.

In a quantum system it is well known that a commuting operator product can be defined
via the time ordering of operators. A definition of a correlation function based on the
time ordered product of two operators has the same commutative properties as the classical
correlation function. The reason why practical quantum mechanics uses noncommuting
products like $\hat{Q}(t_1)\hat{P}(t_2)$ rather than the time ordered product
$T(\hat{Q}(t_1)\hat{P}(t_2))$ is rooted in the subtleties of the definition
of the latter when $t_2=t_1$. Operators for continuous time are defined by a limit process
starting from discrete time steps. The expectation value of the quantum product
$\langle\hat{Q}(t_1)\hat{P}(t_2)\rangle$ is insensitive to the precise definition of the
limiting procedure whereas $\langle T(\hat{Q}(t_1)\hat{P}(t_2))\rangle$ is not.
The ``quantum correlation'' is therefore more ``robust'' that the ``classical correlation''
$\langle T(\hat{Q}(t_1)\hat{P}(t_2))\rangle$. We will see that a similar issue
of a robust definition of correlation functions is actually present in classical statistics
as well. An investigation of the question of relevant information in the classical
probability distribution will lead us to the proposal of a robust quantum correlation for a
classical statistical system.

The formulation of the basic partition function for classical
statistical systems with infinitely many degrees of freedom uses
implicitly an assumption of ``completeness of the statistical
information''. This means that we assign a probability to everyone
of the infinitely many configurations. The specification of the
probability distribution contains therefore an ``infinite amount of
information''. This contrasts with the simple observation that only a
finite amount of information is available in practice for the
computation of the outcome of any physical measurement.
A concentration on measurable quantities suggests that the assumption
of completeness of the statistical information may have to be abandoned.
In this note we explore consequences of ``incomplete statistics'' which deals
with situations where only partial information
about the probability distribution is available. In particular, we
consider extended systems for which only local information about   the
probability distribution is given. From another point of view we ask
which part of the information contained in the standard classical probability distribution
is actually relevant for the computation of expectation values of local observables.
We will see that the quantum mechanical concepts of states, operators and
evolution also emerge naturally in this setting \cite{CWQ}.

As an example, let us consider a classical statistical system where the
infinitely many degrees of freedom $\varphi_n\ (n\in Z)$ are
ordered  in an infinite   chain. We concentrate on a ``local region''
$|\tilde n|<\bar n$ and  assume   that the probability distribution
$p[\varphi]$ has a ``locality property''
in the sense that the relative probabilities
for any two configurations of the ``local
variables'' $\varphi_{\tilde n}$ are independent of the values that
take the  ``external variables''
$\varphi_m$ with $|m|>\bar n$. Furthermore, we
assume that the probability distribution for the $\varphi_{\tilde n}$
is known for given values of the variables $\varphi_{\bar n},
\varphi_{-\bar n}$ at the border of the local interval.
As an example, we may consider a probability distribution
\begin{eqnarray}\label{1.1}
p[\varphi]&=&p_>[\varphi_{m\geq\bar n}]p_0[\varphi_{-\bar n\leq\tilde n
\leq\bar n}]p_<[\varphi
_{m\leq-\bar n}]\nonumber\\
p_0[\varphi]&=&\exp\Bigg\{-\sum_{|\tilde n|<\bar n}\left[\frac{\epsilon}{2}\mu^2
\varphi^2_{\tilde n}+\frac{\epsilon}{8}\lambda\varphi^4_{\tilde n}
+\frac{M}{2\epsilon}(\varphi_{\tilde n}-\varphi_{\tilde n-1})^2\right]\nonumber\\
&&-\frac{M}{2\epsilon}(\varphi_{\bar n}-\varphi_{\bar n-1})^2\Bigg\}
\end{eqnarray}
where $p_>$ and $ p_<$ are only constrained by the overall normalization
of $p[\varphi]$ and we will consider the limit $\epsilon\to 0$.
This statistical system cannot be reduced to a system with a finite number
of degrees of freedom since the probability for the occurrence of
specific values of the ``border variables'' $\varphi_{\bar
n},\varphi_{-\bar n}$ depends on the values of the external
variables  $\varphi_m$ and their probability distribution. The
statistical information about this system is incomplete, if we do
not specify the probability distribution $p_>p_<$ for the external variables
$\varphi_m$ completely.

Local observables are constructed from the local variables
$\varphi_{\tilde n}$. As usual, their expectation values are computed
by ``functional integrals'' where the probability distribution
$p[\varphi]$ appears as a weight factor.
We will ask the question what is the
minimal amount of information about the probability distribution for
the external variables $\varphi_m$
which is necessary for a computation of
expectation values of local observables. One finds that this
information can be summarized in ``states'' $|\psi\},\{\psi|$
that can be represented as ordinary functions $\{\psi(\varphi_{\bar n})|$,
$|\psi(\varphi_{-\bar n})\}$. Since these functions depend each only
on one variable, the
specification of the states contains much less information than
the full probability  distribution $p_>p_<$ which depends on
infinitely many variables $\varphi_{m\geq\bar n}$, $\varphi_{m\leq-\bar n}$.
The states
contain the minimal information for ``local questions'' and  are
therefore the
appropriate quantities for our formulation of incomplete
statistics.
We will see in sect. 6 that any further information about the probability
distributions $p_>[\varphi_m],p_<[\varphi_m]$ beyond the one contained in the
state vectors is actually irrelevant for the computation of expectation values
of local observables.

The expectation values of all local observables can be
computed from the know\-ledge of the local probability distribution and
the states $|\psi\}$ and $\{\psi|$. For this computation one
associates to every local observable $A[\varphi]$ an appropriate operator
$\hat A$ and finds the prescription familiar from quantum mechanics
\begin{equation}\label{1.2}
\langle A[\varphi]\rangle =\{\psi|\hat A|\psi\}\end{equation}
There is a unique mapping $A[\varphi]\to\hat A$ for every
local observable which can be expressed in terms
of an appropriate functional integral. We find that for simple observables $A[\varphi]$ the
operators $\hat A$ correspond precisely to familiar operators in quantum
mechanics. For example, the observable $\varphi(\tilde n)$
can be associated to the operator $\hat Q(\tau)$ in the Heisenberg
picture where time is analytically continued, $\tau=it$, and $\tilde n
=\tau/\epsilon$.

Local correlation functions involving derivatives may be ambiguous in
the continuum limit $\epsilon\rightarrow 0$. This problem is well known in functional
integral formulations of quantum field theories. We show how to
avoid this problem by defining correlations in terms of equivalence
classes of observables. In fact, two
observables $A_1[\varphi],A_2[\varphi]$ can sometimes be represented
by the same operator $\hat A$. In this case $A_1[\varphi]$
and $A_2[\varphi]$ are equivalent since they cannot be
distinguished by their expectation values for arbitrary states.
They have the same expectation values for all possible probability distributions.
We argue that the concept of correlation functions should be based on the
equivalence classes of observables rather than on specific implementations.
Equivalent observables should lead to equivalent correlations. For this
purpose we define a product between equivalence classes of
observables which can be associated to the  product of
operators. For example, we associate a non-commutative product
$\varphi(\tilde n_1)\circ \varphi(
\tilde n_2)$ to the operator product $\hat Q(\tau_1)\hat Q(\tau
_2)$. It is striking how the non-commutativity
of quantum mechanics arises directly from the question what
are meaningful correlation functions. We find that the ``quantum correlation''
based on $\varphi_1\circ\varphi_2$
has better ``robustness properties'' as compared to the usual classical correlation.
We hope that these considerations shed new light on the question if quantum mechanics
can find a formulation in terms of classical statistics \cite{EPR} or general
statistics \cite{GS}.

\bigskip
\section{States and operators}

\medskip
Consider a discrete ordered set of continuous variables
$\varphi_n\equiv\varphi(\tau)$, $\tau=\epsilon n$, $n\in Z$ and a
normalized probability distribution $p(\{\varphi_n\})\equiv
p[\varphi]=\exp(-S[\varphi])$ with $\int D\varphi
e^{-S[\varphi]}\equiv \prod_n(\int
^\infty_{-\infty}d\varphi_n)p[\varphi]=1$.  We will assume that the
action $S$ is local in a range $-\bar\tau<\tau<\bar\tau$, i.e.
\begin{eqnarray}\label{2.1}
S&=&-\ln p=\int^{\bar\tau}_{-\bar\tau}d\tau'{\cal{L}}(\tau')
+S_>(\bar\tau)+S_<(-\bar\tau)\nonumber\\
{\cal{L}}(\tau')&=&V(\varphi(\tau'),\tau')+
\frac{1}{2}Z(\tau')(\partial_{\tau'}\varphi(\tau'))^2\end{eqnarray}
Here we have used a continuum notation $(n_{1,2}=\tau_{1,2}/\epsilon)$
which can be translated into a discrete language by
\begin{eqnarray}\label{2.2}
&&\int^{\tau_2}_{\tau_1}d\tau'{\cal{L}}(\tau')=\epsilon\sum^{n_2-1}
_{n=n_1+1}{\cal{L}}_n+\frac{\epsilon}{2}\left[
V_{n_2}(\varphi_{n_2})+V_{n_1}(\varphi_{n_1})\right]\nonumber\\
&&+\frac{\epsilon}{4}\left[Z_{n_2}\left(\frac{\varphi_{n_2}
-\varphi_{n_2-1}}
{\epsilon}\right)^2+Z_{n_1}\left(\frac{\varphi_{n_1+1}
-\varphi_{n_1}}{\epsilon}\right)^2\right] \end{eqnarray}
with
\begin{equation}\label{2.2a}
{\cal L}_n=V_n(\varphi_n)+\frac{Z_n}{4\epsilon^2}\{(\varphi_{n+1}
-\varphi_n)^2+(\varphi_n-\varphi_{n-1})^2\}\end{equation}
This corresponds to a discrete derivative
\begin{eqnarray}\label{2.3}
(\partial_\tau\varphi(\tau))^2&=&\frac{1}{2}
\left\{\left(\frac{\varphi
(\tau+\epsilon)-\varphi(\tau)}{\epsilon}\right)^2+ \left(\frac{\varphi
(\tau)-\varphi(\tau-\epsilon)}{\epsilon}\right)^2\right\}\nonumber\\
&=&\frac{1}{2\epsilon^2}\left\{\left(\varphi_{n+1}-\varphi_n\right)^2
+\left(\varphi_n-\varphi_{n-1}\right)^2\right\}.\end{eqnarray}
The boundary
terms in (\ref{2.2}) are chosen such that $S_>(\bar\tau)$ is
independent of all $\varphi(\tau')$ with $\tau'<\bar\tau$ whereas
$S_<(-\bar\tau)$ only depends on $\varphi(\tau'\leq-\bar\tau)$. Except
for the overall normalization of $p$ no additional assumptions about
the form of $S_>(\bar\tau)$ and $S_<(-\bar\tau)$ will be made. In case
of $S$ being local also at $\bar\tau$ we note that $S_>(\bar\tau)$
contains a term $\frac{\epsilon}{2}[V(\varphi(\bar\tau),\bar\tau)+
V(\varphi(\bar\tau+\epsilon),\bar\tau+\epsilon)]+\frac{\epsilon}
{4}(Z(\bar\tau)+Z(\bar\tau+\epsilon))\left(\frac{
\varphi(\bar\tau+\epsilon)-\varphi(\bar\tau)}{\epsilon}\right)^2$,
which involves a product $\varphi(\bar\tau+\epsilon)\varphi(\bar\tau)$
and therefore links the variables with $\tau>\bar\tau$ to the ones
with $\tau\leq \bar\tau$.

We are interested in local observables $A[\varphi;\tau]$ which depend
only on those $\varphi(\tau')$ where $\tau-\frac{\delta}
{2}\leq\tau'\leq\tau+\frac{\delta}{2}$.  (We assume
$-\bar\tau<\tau-\frac{\delta}{2}$, $\bar\tau> \tau+\frac{\delta}{2}$.)
As usual, the expectation value of $A$ is
\begin{equation}\label{2.4}
<A(\tau)>=\int
D\varphi A[\varphi;\tau]e^{-S[\varphi]} \end{equation}
As mentioned in the introduction, our
investigation concerns the question what we can learn about
expectation values of local observables and suitable   products
thereof in a situation where we have no or only partial information
about $S_>(\bar\tau)$ and $S_<(-\bar\tau)$.  It seems obvious that the
full information contained in $S$ is not needed if only expectation
values of local observables of the type (\ref{2.4}) are to be computed.
On the
other hand, $<A(\tau)>$ cannot be completely independent of $S_>
(\bar\tau)$ and $S_<(-\bar\tau)$ since the next neighbor interactions
(\ref{2.2}) relate ``local variables'' $\varphi(\tau-\frac{\delta}{2}<\tau'
<\tau+\frac{\delta}{2})$ to the ``exterior variables'' $\varphi(\tau'>\bar\tau)$ and
$\varphi(\tau'<-\bar\tau)$.

In order to establish the necessary amount of information needed from
$S_>(\bar\tau)$ and $S_<(-\bar\tau)$ we first extend $S_>$ and $S_<$
to values $|\tau|<\bar\tau$
\begin{equation}\label{2.5a}
S_<(\tau_1)=S_<(-\bar\tau)+\int^{\tau_1}
_{-\bar\tau}d\tau'{\cal{L}}(\tau')\ ,\
 S_>(\tau_2)=S_>(\bar\tau)+\int^{\bar\tau}_{\tau_2}
d\tau'{\cal{L}}(\tau')\end{equation}
where we note the general identity
\begin{equation}\label{2.5}
S_>(\tau)+S_<(\tau)=S \end{equation}
The expectation value (\ref{2.4}) can be written as
\begin{eqnarray}\label{2.6}
<A(\tau)>&=&\int
d\varphi(\tau+\frac{\delta}{2})\int
d\varphi(\tau-\frac{\delta}{2})\int
D\varphi_{(\tau'>\tau+\frac{\delta}
{2})}e^{-S_>(\tau+\frac{\delta}{2})}\nonumber\\
&&\int
D\varphi_{(\tau-\frac{\delta}{2}<\tau'<\tau+\frac{\delta}{2})}
A[\varphi;\tau]\exp\{-\int^{\tau+\frac{\delta}{2}}
_{\tau-\frac{\delta}{2}}d\tau''{\cal{L}}(\tau'')\}\nonumber\\
&&\int D\varphi_{(\tau'<\tau-\frac{\delta}{2})}e^{-S_<(\tau-
\frac{\delta}{2})}
\end{eqnarray}
This suggests the introduction of the
``states''
\begin{eqnarray}\label{2.7}
&&|\psi(\varphi(\tau-\frac{\delta}{2});\
\tau-\frac{\delta}{2})\} =\int
D\varphi_{(\tau'<\tau-\frac{\delta}{2})}e^{-S_<(\tau
-\frac{\delta}{2})}\nonumber\\
&&\{\psi(\varphi(\tau+\frac{\delta}{2});\ \tau+\frac{\delta}{2})|
=\int D\varphi_{(\tau'>\tau+\frac{\delta}{2})}
e^{-S_>(\tau+\frac{\delta}{2})}\end{eqnarray}
and the operator
\begin{eqnarray}\label{2.8}
&&\hat A_\delta(\varphi(\tau+\frac{\delta}{2}),\ \varphi(\tau-
\frac{\delta}{2});\ \tau)\nonumber\\ &=&\int
D\varphi_{(\tau-\frac{\delta}{2}<\tau'<\tau+\frac{\delta}{2})}
A[\varphi;\ \tau]\exp\left\{-\int^{\tau+\frac{\delta}{2}}
_{\tau-\frac{\delta}{2}}d\tau''{\cal{L}}(\tau'')\right\}\end{eqnarray}
We note that
$|\psi\}$ is a function of $\varphi(\tau- \frac{\delta}{2})$ since the
latter appears in $S_<(\tau-\frac{\delta} {2})$ and is not included in
the (``functional'') integration (\ref{2.7}). Similarly, $\{\psi|$
depends on $\varphi(\tau +\frac{\delta}{2})$ whereas $\hat A$ is a
function of the two variables $\varphi(\tau +\frac{\delta}{2})$ and
$\varphi(\tau-\frac{\delta}{2})$. Using a notation where $|\psi\}$ and
$\{\psi|$ are interpreted as (infinite dimensional) vectors and $\hat
A$ as a matrix, one has
\begin{eqnarray}\label{2.9}
&&<A(\tau)>=\{\psi(\tau+\frac{\delta}{2})\hat A_\delta
(\tau)\psi(\tau-\frac{\delta}{2})\}\\ &&\equiv\int
d\varphi_2\int d\varphi_1\{\psi(\varphi_2;\tau+\frac{\delta} {2})|\hat
A_\delta(\varphi_2,\varphi_1;\tau)|
\psi(\varphi_1;\tau-\frac{\delta}{2})\}\nonumber\end{eqnarray}
This form resembles
already the well-known prescription for expectation values of
operators in quantum mechanics.  In contrast to quantum mechanics
(\ref{2.9}) still involves, however, two different state vectors.

The mapping $A[\varphi;\ \tau]\to\hat A_\delta(\tau)$ can be computed
(cf. (\ref{2.8})) if ${\cal{L}}(\tau')$ is known for
$|\tau'|<\bar\tau$. The only information needed from $S_>(\bar\tau)$
and $S_<(-\bar\tau)$ is therefore contained in the two functions
$\{\psi(\varphi)|$ and $|\psi(\varphi)\}$! The specification of these
states (wave functions) at $\bar\tau$ and $-\bar\tau$ and of ${\cal{L}}
(|\tau|<\bar\tau)$ completely determines the expectation values of
{\em all} local observables!

We will see below the close
connection to the states in quantum mechanics. In our context we
emphasize that for any given $S$ these states can be computed as well
defined functional integrals (\ref{2.7}). Due to (\ref{2.5}) they obey
the normalization
\begin{equation}\label{2.10} \{\psi(\tau)\psi(\tau)\} \equiv\int
d\varphi\{\psi(\varphi;\tau)||\psi(\varphi;\tau)\}=1\end{equation}
Incomplete statistics explores statements that can be made
for local observables and appropriate products thereof  without using information about $S_>$ or $S_<$ beyond the
one contained in the states $|\psi\}$ and $\{\psi|$.

\bigskip
\section{ Evolution in Euclidean time}

\medskip
For a ``locality interval'' $\delta>0$ the expression (\ref{2.9})) involves
states at different locations or
``Euclidean times'' $\tau+\frac{\delta} {2}$ and
$\tau-\frac{\delta}{2}$. We aim for a formulation where only states at
the same $\tau$ appear. We therefore need the explicit mapping from
$|\psi(\tau-\frac{\delta}{2})\}$ to a reference point $|\psi(\tau)\}$
and similar for $\{\psi(\tau+\frac{\delta}{2})|$. This mapping
should also map $\hat A_\delta$ to a suitable operator such that the
structure (\ref{2.9}) remains preserved.
The dependence of states and operators on the
Euclidean time $\tau$ is described by evolution operators
$(\tau_2>\tau_1,\ \tau_2>\tau_f,\
\tau_i>\tau_1,\tau_f=\tau+\frac{\delta}{2},\
\tau_i=\tau-\frac{\delta}{2}))$
\begin{eqnarray}\label{3.1}
&&|\psi(\tau_2)\}=\hat
U(\tau_2,\tau_1)|\psi(\tau_1)\}\nonumber\\
&&\{\psi(\tau_1)|=\{\psi(\tau_2)|\hat U(\tau_2,\tau_1)\nonumber\\
&&\hat A(\tau_2,\tau_1)=\hat U(\tau_2,\tau_f)\hat A
(\tau_f,\tau_i)\hat U(\tau_i,\tau_1) \end{eqnarray}
or differential operator equations $(\epsilon\to0)$
\begin{equation}\label{3.2}
\partial_\tau|\psi(\tau)\}=-\hat H(\tau)|\psi(\tau)\}
\end{equation}
The evolution operator has an explicit
representation as a functional integral
\begin{equation}\label{3.3} \hat
U(\varphi(\tau_2),\varphi(\tau_1);\ \tau_2,\tau_1) =\int
D\varphi_{(\tau_1<\tau'<\tau_2)}\exp\left\{
-\int^{\tau_2}_{\tau_1}d\tau''{\cal{L}}(\tau'')\right\}\end{equation}
and obeys the
composition property $(\tau_3>\tau_2>\tau_1)$
\begin{equation}\label{3.4} \hat
U(\tau_3,\tau_2)\hat U(\tau_2,\tau_1)=\hat U(\tau_3,\tau_1)\end{equation}
with
\begin{equation}\label{3.5} \hat U(\varphi_2,\varphi_1;
\tau,\tau)=\delta(\varphi_2-\varphi_1)\end{equation}
It can therefore be composed
as a product of transfer matrices or ``infinitesimal'' evolution
operators
\begin{equation}\label{3.6}
\hat U(\tau+\epsilon,\tau)=e^{-\epsilon\hat
H(\tau+\frac{\epsilon}{2})}\end{equation}

In case of translation symmetry for the local part of the probability
distribution, i. e. for $V$ and $Z$ independent of $\tau$,  we note the
symmetry in $\varphi_1\leftrightarrow\varphi_2$
\begin{equation}\label{3.7} \hat
U(\tau+\epsilon,\tau)=\hat U^T(\tau+\epsilon,\tau)\ ,\ \hat H(\tau
+\frac{\epsilon}{2})= \hat H^T(\tau+\frac{\epsilon}{2})=\hat H\end{equation}
In this case the real symmetric matrix $\hat H$ has real eigenvalues
$E_n$. Then the general solution of the differential equation
(\ref{3.2}) may be written in the form
\begin{equation}\label{3.8}
|\psi(\tau)\}=\sum_n\psi_0^{(n)}e^{-E_n\tau},\
\{\psi(\tau)|=\sum_n\bar\psi_0^{(n)}e^{E_n\tau}\end{equation}
where $\psi_0^{(n)}$ and $\bar\psi_0^{(n)}$ are eigenvectors
of $\hat H$ with eigenvalues $E_n$. Here we recall that the construction
(\ref{2.7}) implies that $|\psi\}$ and $\{\psi|$ are real positive functions
of $\varphi$ for every $\tau$. This restricts the allowed values of the
coefficients $\psi^{(n)}_0,\bar{\psi}^{(n)}_0$.

We next want to compute the explicit form of the Hamilton
operator $\hat H$. It is fixed uniquely by the functional
integral representation (\ref{3.3}) for $\hat U$. In order to
obey the defining equation (\ref{3.6}), the  Hamilton
operator $\hat H$ must fulfill for arbitrary $|\psi(\varphi)\}$ the
relation (with
$Z=Z(\tau+\frac{\epsilon}{2})=\frac{1}{2}(Z(\tau+\epsilon) +Z(\tau))$
\begin{eqnarray}\label{3.9}
&&\int d\varphi_1\hat
H(\varphi_2,\varphi_1)|\psi(\varphi_1)\}
=-\lim_{\epsilon\to0}\frac{1}{\epsilon}\Big\{ \int d\varphi_1\\
&&\exp\left[-\frac{\epsilon}{2}(V(\varphi_2)
+V(\varphi_1))-\frac{Z}{2\epsilon}(\varphi_2-\varphi_1)^2\right]
|\psi(\varphi_1)\}-|\psi(\varphi_2)\}\Big\}\nonumber
\end{eqnarray}
The solution of this equation can be expressed in terms of the operators
\begin{eqnarray}\label{3.10}
&&\hat Q(\varphi_2,\varphi_1)=\varphi_1\delta(\varphi_2-\varphi_1)
\nonumber\\ &&\hat
P^2(\varphi_2,\varphi_1)=-\delta(\varphi_2-\varphi_1)
\frac{\partial^2}{\partial \varphi_1^2}\end{eqnarray}
as
\begin{equation}\label{3.11}
\hat H(\tau)=V(\hat Q,\tau)+\frac{1}{2Z(\tau)} \hat P^2\end{equation}
This can be established by using under the
$\varphi_1$-integral the replacement
\begin{equation}\label{3.12}
e^{-\frac{Z}{2\epsilon}(\varphi_2-\varphi_1)^2}
\to\left(\frac{2\pi\epsilon}{Z}\right)^{1/2}\delta
(\varphi_2-\varphi_1)\exp\left(\frac{\epsilon}{2Z}\frac{\partial^2}
{\partial\varphi^2_1}\right)\end{equation}
which is valid by partial integration
if the integrand decays fast enough for $|\varphi_1|\to\infty$. We
note that the operators $\hat Q$ and $\hat P^2$ do not commute, e.g.
\begin{equation}\label{3.13} [\hat P^2,\hat
Q](\varphi_2,\varphi_1)=-2\delta(\varphi_2-\varphi_1)
\frac{\partial}{\partial\varphi_1}\end{equation}

The Hamilton operator can be used in order to establish the existence
of the inverse of the ``infinitesimal'' evolution operator, $\hat
U^{-1}(\tau+\epsilon,\tau)=e^{\epsilon\hat H
(\tau+\frac{\epsilon}{2})}$. Then the inverse $\hat U^{-1}
(\tau_2,\tau_1)$ is defined by the multiplication of ``infinitesimal''
inverse evolution operators, and we can extend the composition
property (\ref{3.4}) to arbitrary $\tau$ be defining for
$\tau_2<\tau_1$
\begin{equation}\label{3.14} \hat U(\tau_2,\tau_1)=\hat
U^{-1}(\tau_1,\tau_2)\end{equation}
(For a given dependence of $\hat U$ on the
variables $\tau_2$ and $\tau_1$ the matrix $\hat U(\tau_1,\tau_2)$
obtains from $\hat U(\tau_2,\tau_1)$ by a simple exchange of the
arguments $\tau_1$ and $\tau_2$.) Using (\ref{3.1}), this allows us to
write the expectation value of a local observable in a form involving
states at the same $\tau$-variable
\begin{equation}\label{3.15}
<A(\tau)>=\{\psi(\tau)\hat U(\tau,\tau+\frac{\delta}{2}
)\hat A_\delta(\tau)\hat U(\tau-\frac{\delta}{2},\tau)\psi(\tau)\}\end{equation}

\bigskip
\section { Schr\"odinger and Heisenberg operators}

\medskip
In this section we want to exploit further the mapping between
incomplete statistics and quantum mechanics for situations where
expectation values like $<\varphi(\tau)>$ may depend on $\tau$. A typical
question one may ask within incomplete classical statistics is the
following: Given a large set of measurements of observables with support
at a given value $\tau=0$, like $<\varphi^p(0)>, <(\partial_\tau\varphi(0))
^{p'}>$, etc., what can one predict for the expectation values of similar
observables at some other location $\tau\not=0$? It is obvious
that the evolution operator $\hat U$ is the appropriate tool to tackle
this type of questions.

The existence of the inverse evolution operator allows us to associate
to an observable $A(\tau)$ the operator $\hat A_S(\tau)$ in the
Schr\"odinger representation (cf. (\ref{3.15}))
\begin{equation}\label{4.1}
\hat A_S(\tau)=\hat
U(\tau,\tau+\frac{\delta}{2})\hat A_\delta (\tau)\hat
U(\tau-\frac{\delta}{2},\tau)\end{equation}
The expectation value of the
observable $A$ can be expressed by the expectation value of the
operator $\hat A_S$ in a way analogous to quantum mechanics
\begin{equation}\label{4.2} <A(\tau)>=\{\psi(\tau)|\hat A_S(\tau)
|\psi(\tau)\} ={\rm
Tr}\rho(\tau)\hat A_S(\tau)\end{equation}
For the second identity we have
introduced the ``density matrix''
\begin{eqnarray}\label{4.3}
\rho(\varphi_1,\varphi_2,\tau)&=&|\psi(\varphi_1,\tau\}\{\psi(\varphi
_2,\tau)|=\int D\varphi_{(\tau'\not=\tau)}e^{-S(\varphi_1,\varphi_2)}
\nonumber\\ {\rm Tr}\rho(\tau)&=&1\end{eqnarray}
where $S(\varphi_1,\varphi_2)$
obtains from $S$ by replacing $\varphi(\tau)\to\varphi_1$ for all
``kinetic'' terms involving $\varphi(\tau'<\tau)$ and
$\varphi(\tau)\to\varphi_2$ for those involving $\varphi(\tau'>\tau)$,
whereas for potential terms $e^{-\epsilon V(\varphi(\tau))}\to
e^{-\frac{\epsilon}{2}(V(\varphi_1)+V(\varphi_2))}$.

In order to make the transition to the Heisenberg picture, we
may select a reference point $\tau=0$ and define
\begin{equation}\label{4.4} \hat
U(\tau)\equiv \hat U(\tau,0)\ ,\ \rho\equiv\rho(\tau=0)\ ,\
\rho(\tau)=\hat U(\tau)\rho\hat U^{-1}(\tau)\end{equation}
This specifies the
Heisenberg picture for the $\tau$-dependent operators
\begin{eqnarray}\label{4.5}
\hat A_H(\tau)&=&\hat U^{-1}(\tau)\hat A_S(\tau)\hat
U(\tau)\nonumber\\ <A(\tau)>&=&{\rm Tr}\rho \hat A_H(\tau)\end{eqnarray}
We note
that for two local observables $A_1, A_2$ the linear combinations
$A=\alpha_1A_1+\alpha_2A_2$ are also local observables. The associated
operators obey the same linear relations $\hat A=\alpha_1\hat
A_1+\alpha_2\hat A_2$, where $\hat A$ stands for $\hat A_\delta,\hat
A_S$ or $\hat A_H$. The relation (\ref{4.5}) is the appropriate
formula to answer the question at the beginning of this section.
One may use the set of measurements
of expectation values at $\tau=0$ to gather information about $\rho$.
Once $\rho$ is determined with sufficient accuracy, the expectation
values $<A(\tau)>$ can be computed. Of course, this needs a computation of
the explicit form of the Heisenberg operator $\hat A_H(\tau)$.

It is instructive to observe that some simple local observables have a
$\tau$-independent operator representation in the
Schr\"odinger picture. This is easily seen for observables $A(\tau)$
which depend only on the variable $\varphi(\tau)$. The mapping reads
\begin{equation}\label{4.6}
A(\tau)=f(\varphi(\tau))\to\hat A_S(\tau)=f(\hat Q)\end{equation}
Observables depending only on one variable $\varphi(\tau)$ therefore have the
Heisenberg representation (cf. (\ref{4.6}))
\begin{equation}\label{4.11}
A(\tau)=f(\varphi(\tau))\to \hat A_H(\tau)=
f(\hat Q(\tau))\end{equation}
Here we have used the definition
\begin{equation}\label{4.12}
\hat Q(\tau)=\hat U^{-1}(\tau)\hat Q\hat U(\tau)\end{equation}
More generally, one finds for products of functions depending on the
variables $\varphi(\tau_1),\varphi(\tau_2)...\varphi(\tau _n)$ with
$\tau_1<\tau_2<...\tau_n$ the Heisenberg operator
\begin{eqnarray}\label{4.13}
A(\tau_1,...\tau_n)&=&f_1(\varphi(\tau_1)f_2(\varphi(\tau_2))...
f_n(\varphi(\tau_n))\longrightarrow\nonumber\\ \hat A_H(\tau)&=&\hat
U^{-1}(\tau_n)f_n(\hat Q)\hat U(\tau_n, \tau_{n-1})...
\hat U(\tau_2,\tau_1) f_1(\hat Q)\hat
U(\tau_1)\nonumber\\ &=&f_n(\hat Q(\tau_n))...f_2(\hat
Q(\tau_2))f_1(\hat Q(\tau_1))\end{eqnarray}
This important relation follows
directly from the definitions (\ref{2.8}), (\ref{4.1}), (\ref{4.5}). We
observe that $\hat A_H$ depends on the variables $\tau_i$ which are
the arguments of $A$ but shows no dependence on the reference point
$\tau$. (Only $\hat A_\delta$ and $\hat A_S$ depend on $\tau$.)

We can
use (\ref{4.13}) to find easily the Heisenberg operators for
observables involving ``derivatives'', e.g.
\begin{eqnarray}\label{4.14}
A&=&\tilde\partial_\tau\varphi(\tau_1)=\frac{1}{2\epsilon}
(\varphi(\tau_1+\epsilon)-\varphi(\tau_1-\epsilon))\nonumber\\
\hat
A_H&=&\frac{1}{2\epsilon}\{\hat U^{-1}(\tau_1+\epsilon)\hat Q\hat U(
\tau_1+\epsilon)-\hat U^{-1}(\tau_1-\epsilon)\hat Q\hat
U(\tau_1-\epsilon)\} \nonumber\\
&=&-\frac{1}{Z(\tau_1)}\hat R(\tau_1)+O(\epsilon)\end{eqnarray}
where we have assumed that $\hat H$ is a
smooth function of $\tau$. Here $\hat R$ is defined by
\begin{equation}\label{4.15} \hat
R(\varphi_2,\varphi_1)=\delta(\varphi_2-\varphi_1)\frac{\partial}
{\partial\varphi_1} \ ,\  \hat R^2=-\hat P^2\end{equation}
and we use, similar to (\ref{4.12}), the definitions
\begin{equation}\label{5.15A}
\hat R(\tau)=\hat U^{-1}(\tau)\hat R\hat U(\tau)\ ,\quad \hat P
^2(\tau)=\hat U^{-1}(\tau)\hat P^2\hat U(\tau)\end{equation}

Two different definitions of derivatives can lead to the same
operator $\hat A_H$. An example is the observable
\begin{equation}\label{4.15a}
A=\partial^>_\tau\varphi(\tau_1)=\frac{1}{\epsilon}(\varphi(\tau_1+\epsilon)
-\varphi(\tau_1))\end{equation}
Up to terms of order $\epsilon$ the associated Heisenberg operator
is again given by $\hat A_H=-Z(\tau_1)^{-1}
\hat U^{-1}(\tau_1)\hat R\hat U(\tau_1)$
and therefore the same as for $\tilde\partial_\tau\varphi(\tau_1)$
(\ref{4.14}). Applying the
same procedure to the squared derivative observable (\ref{2.3}) yields
\begin{equation}\label{4.16}
A=(\partial_\tau\varphi)^2(\tau_1)\longrightarrow\hat A_H=
\frac{1}{\epsilon Z}-\frac{1}{Z^2}\hat P^2(\tau_1)
\end{equation}
where we have assumed for
simplicity a $\tau$-independent Hamiltonian $\hat H$. It is remarkable that this operator
differs from the square of the Heisenberg operator associated
to $\tilde\partial_\tau\varphi(\tau_1)$ by a constant which diverges for $\epsilon\to0$.
Indeed, one finds
\begin{eqnarray}\label{5.17}
A&=&(\tilde\partial_\tau\varphi(\tau_1))^2\to\hat A_H=\frac{1}{2\epsilon Z}-
\frac{1}{Z^2}\hat P^2(\tau_1)\nonumber\\
A&=&(\partial^>_\tau\varphi(\tau_1))^2\to \hat A_H=\frac{1}{\epsilon Z}-\frac{1}{Z^2}\hat P^2(\tau_1)
\end{eqnarray}
Equation (\ref{5.17}) teaches us that the product
of derivative observables with other observables can be
ambiguous in the sense that the associated operator and expectation
value depends very sensitively on the precise definition of the
derivative.
This ambiguity of
the derivative observables in the continuum limit is an unpleasant
feature for the formulation of correlation functions. It survives when the discussion
is extended to observables that are smoothened over a certain interval instead of
being strictly local \cite{CWQ}.

In the next sections
we will see how this problem is connected with the concept of quantum
correlations. We will argue that the ambiguity in the classical
correlation may be the basic ingredient why a description of our
world in terms of quantum statistics is superior to the use of
classical correlation functions.

\bigskip
\section{Correlation functions}

A basic concept for any statistical description are correlation functions
for a number of observables $A_1[\varphi], A_2[\varphi], ...$ In particular,
a two-point function is given by
the expectation value of an associative product of two observables
$A_1[\varphi]$ and $A_2[\varphi]$. For local observables $A_1,A_2$ the
product should again be a local observable which must be defined
uniquely in terms of the definitions of $A_1$ and $A_2$. This
requirement, however, does not fix the definition of the correlation
uniquely. The standard ``classical product'', i.e. the simple
multiplication of the functionals $A_1[\varphi]\cdot A_2[\varphi]$
(in the same sense as the ``pointwise'' multiplication of functions)
fulfills the general requirements\footnote{This holds provided that
the product results in a meaningful observable with finite expectation
value.} for a correlation function. Other
definitions can be conceived as well. In this section we will introduce
a quantum correlation which equals the classical (``pointwise'')
correlation only for $\tau$-ordered non-overlapping observables.
In contrast, for two local observables with overlapping support
we will find important differences between the quantum and classical
correlation. In particular, we will discover the
effects of the non-commutativity characteristic for quantum mechanics.

Incomplete statistics draws our attention to
an important issue in the formulation of meaningful correlation functions.
 Consider the two versions
of the derivative observable $\tilde\partial_\tau\varphi$ and
$\partial^>_\tau\varphi$ defined by eqs. (\ref{4.14}) and
(\ref{4.15}), respectively. In the continuum limit $(\epsilon\to 0)$ they
are represented by the same operator $\hat A_H$. In consequence,
both definitions lead to the same expectation value for any state
$|\psi\},\{\psi|$.  The two
versions of derivative observables cannot be distinguished
by any measurement and should
therefore be identified. On the other hand, the classical products
$\tilde\partial_\tau\varphi(\tau_1)\cdot\tilde\partial_\tau
\varphi(\tau_2)$ and $\partial^>_\tau\varphi(\tau_1)\cdot\partial
_\tau^>\varphi(\tau_2)$ are represented by different operators
for $\tau_1=\tau_2$,
as can be seen from (\ref{5.17}).
This means that  the two versions of derivative
observables lead to different classical correlation functions! Obviously,
this situation is unsatisfactory since for $\epsilon\to 0$
no difference between the two versions could  be ``measured''
for the observables themselves. We find this disease
unacceptable for a meaningful correlation and require
as a criterion for a meaningful correlation
function  that two observables which have the same
expectation values for all (arbitrary) probability
distributions should also have identical correlation functions.
We have shown that two observables
which are represented by the same Heisenberg operator have indeed
the same expectation values for all possible probability
distributions and should therefore be considered as equivalent.
They should therefore
lead to indistinguishable correlation functions.

As we have already established, the two derivative observables
$A_1=\tilde\partial_\tau
\varphi(\tau)$ and $A_2=\partial^>_\tau\varphi(\tau)$ are
indistinguishable in the continuum limit, whereas their classical correlations are not.
We may therefore conclude
that the classical correlation $A_1\cdot A_2$ is not a meaningful correlation function.
In this section we propose the use of a different correlation
based on a quantum product $A_1\circ A_2$. By construction,
this correlation will always obey our criterion of ``robustness''
with respect to the precise choice of the observables. It should
therefore be considered as an interesting alternative to
the classical correlation. At this place we only note that the
``robustness problem'' is not necessarily connected to
the continuum limit. The mismatch between
indistinguishable observables and distinguishable ``classical'' correlations
can appear quite generally also for $\epsilon>0$.

Our formulation of a quantum correlation will be based on the concepts
of  equivalent observables and products defined for equivalence
classes. In fact, the
mapping $A(\tau)\to\hat A_H(\tau)$ is not necessarily
invertible on the space
of all observables $A(\tau)$. This follows from the simple observation
that already the map (\ref{2.8}) contains integrations.
Two different integrands (observables)
could lead to the same value of the integral (operator)
for arbitrary  fixed boundary
values $\varphi(\tau-\frac{\delta}{2}),\ \varphi(\tau+\frac{
\delta}{2})$. It is
therefore possible that two different observables $A_a
(\tau)$ and $A_b(\tau)$ can be mapped into the same
Heisenberg operator $\hat
A_H(\tau)$. Since the expectation values can be computed from $\hat
A_H(\tau)$ and $\rho$ only, no distinction between $<A_a>$ and $<A_b>$
can then be made for arbitrary $\rho$. All local observables $A(\tau)$
which correspond to the same operator $\hat A_H(\tau)$ are
equivalent.

We are interested in structures that only
depend on the equivalence classes of observables. Addition of two
observables and multiplication with a scalar can simply be carried
over to the operators. This is not necessarily the case, however, for the
(pointwise) multiplication of two observables.  If $A_a{(\tau)}$ and
$A_b{(\tau)}$ are both mapped into $\hat A_H(\tau)$ and a third observable
$B(\tau)$ corresponds to $\hat B_H(\tau)$, the
products $A_a\cdot B$ and $A_b \cdot B$ may nevertheless be
represented by different operators.
It is then easy to construct states
where $<A_aB> \not=$$<A_bB>$ and the pointwise product does not
depend only on the equivalence class.

On the other
hand, the (matrix) product of two operators $\hat A_H \hat B_H$
obviously refers only to the equivalence class. It can be implemented
on the level of observables by defining a unique ``standard
representative'' of the equivalence class as
\begin{equation}\label{5.1} \bar
A[\varphi,\tau]=F[\hat A_H(\tau)]\end{equation}
Using the mapping $A[\tau]\to\hat
A_H(\tau)$ (\ref{2.8}), (\ref{4.1}), (\ref{4.5}), we define
the quantum product of
two observables as
\begin{equation}\label{5.2}
A(\varphi,\tau)\circ
B(\varphi,\tau)=F[\hat A_H(\tau)\hat B_H(\tau)] \equiv(A\circ
B)[\varphi,\tau]\end{equation}
This product is associative, but not commutative. (By
definition, the operator associated to the observable $(A\circ
B)(\varphi,\tau)$ is $\hat A_H(\tau)\hat B_H(\tau)$ and the product
$A\circ B$ is isomorphic to the ``matrix multiplication'' $\hat A\hat
B$ if restricted to the subspace of operators $\bar A=F[\hat A], \bar
B=F[\hat B]$.)
The correlations (e.g. expectation values of products of
observables) formed with the product $\circ$ reflect the
non-commutative structure of quantum mechanics.
This justifies the name ``quantum correlations''.
Nevertheless, we emphasize that the
``quantum product'' $\circ$ can also be viewed as
just a particular structure among
``classical observables''.

The definition of the quantum product is unique on the level of operators.
On the level of
the classical observables, it is, however, not yet fixed uniquely
by (\ref{5.2}). The precise definition
obviously depends on the choice of a
standard representation $F[\hat A_H(\tau)]$ for the equivalence class of
observables represented by $\hat{A}_H$. We will  choose a linear
map $F[\alpha_1\hat A_{H,1}+\alpha_2\hat A_{H,2}] =\alpha_1F[\hat
A_{H,1}]+\alpha_2 F[\hat A_{H,2}]$ with the property that it inverses
the relation (\ref{4.13}). For ``time-ordered''
$\tau_1<\tau_2<...\tau_n$ the map $F$ should then obey
\begin{equation}\label{5.3}
F[f_n(\hat Q(\tau_n))...f_2(\hat Q(\tau_2))f_1(\hat Q(\tau_1))]
=f_1(\varphi(\tau_1))f_2(\varphi(\tau_2))...f_n(\varphi(\tau_n)).\end{equation}
It is easy to
see how this choice exhibits directly the noncommutative
property of the quantum
product between two observables. As an example let us consider the two
observables $\varphi(\tau_1)$ and $\varphi(\tau_2)$ with
$\tau_1<\tau_2$. The quantum product or quantum correlation
depends on the ordering
\begin{eqnarray}\label{5.4}
\varphi(\tau_2)\circ\varphi(\tau_1)&=&\varphi(\tau_2)\varphi(\tau_1)
\nonumber\\
\varphi(\tau_1)\circ\varphi(\tau_2)&=&\varphi(\tau_2)
\varphi(\tau_1)+F[[\hat Q(\tau_1),\hat Q(\tau_2)]]\end{eqnarray}
The
noncommutative property of the quantum product for these operators is
directly related to the commutator
\begin{eqnarray}\label{5.5} [\hat Q(\tau_1),\
\hat Q(\tau_2)]&=&\hat U^{-1}(\tau_1)\hat Q\hat   U(\tau_1,\tau_2)\hat
Q \hat U(\tau_2)\nonumber\\
&&-\hat U^{-1}(\tau_2)\hat Q\hat
U(\tau_2,\tau_1)\hat Q\hat U(\tau_1)\end{eqnarray}
Only for time-ordered arguments the quantum correlation coincides
with the classical correlation.

The map $F$ can easily be extended to operators involving
derivatives of $\varphi$. We concentrate here for simplicity on
a translation invariant probability distribution in the local
region with constant $Z(\tau)=Z$.
The mappings (with $\tau_2\geq \tau_1+\epsilon)$
\begin{eqnarray}\label{6.6}
F(\hat R(\tau))&=&-Z\partial^>_\tau\varphi(\tau)\nonumber\\
F(\hat R(\tau)\hat Q(\tau))&=&-Z\varphi(\tau)\partial^>_\tau\varphi
(\tau)\nonumber\\
F(\hat R(\tau_2)\hat R(\tau_1))&=&Z^2\partial^>_\tau\varphi(\tau_2)\partial^>_\tau\varphi(
\tau_1)\end{eqnarray}
are compatible with (\ref{5.3}). This can be seen by noting that
the $\tau$-evolution of $\hat Q(\tau)$ according to (\ref{4.12})
implies for $\epsilon\to 0$ the simple relation
\begin{equation}\label{6.7}
\partial_\tau\hat Q(\tau)=[\hat H,\hat Q(\tau)]=-Z^{-1}\hat R(\tau)\end{equation}
A similar construction (note $[\hat Q(\tau+\epsilon), \hat Q(\tau)]
=-\epsilon/Z)$ leads to
\begin{equation}\label{6.8}
F(\hat R^2(\tau))=Z^2(\partial_\tau^>\varphi(\tau))^2-Z/\epsilon\end{equation}
and we infer that the quantum product of derivative observables at
equal sites differs from the pointwise product
\begin{equation}\label{6.9}
\partial^>_\tau\varphi(\tau)\circ\partial^>_\tau\varphi(\tau)=(\partial
^>_\tau\varphi(\tau))^2-1/(\epsilon Z)\end{equation}

From the relations (\ref{5.4}) and (\ref{6.9}) it has become clear
that the difference between the quantum product and the ``pointwise''
classical product of two observables is related to their $\tau$-ordering
and ``overlap''. Let us define that two observables $A_1[\varphi]$
and $A_2[\varphi]$ overlap if they depend on variables $\varphi(\tau)$
lying in two overlapping $\tau$-ranges ${\cal R}_1$ and ${\cal R}_2$.
(Here two
ranges do not overlap if all $\tau$ in ${\cal R}_1$ obey $\tau\leq \tau_0$ whereas
for ${\cal R}_2$ one has $\tau\geq \tau_0$, or vice versa.
This implies that non-overlapping observables can depend on at
most one common variable $\varphi(\tau_0)$.) With this definition
the quantum product is equal to the classical product if the observables
do not overlap and are $\tau$-ordered (in the sense that larger $\tau$
are on the left side).

In conclusion, we have established a one-to-one correspondence between
classical correlations $\varphi(\tau_2)\varphi(\tau_1)$ and the
product of Heisenberg operators $\hat Q(\tau_2)\hat Q(\tau_1)$ provided
that the $\tau$-ordering $\tau_2\geq\tau_1$ is respected. This extends
to observables that can be written as sums or integrals over $\varphi(\tau)$
(as, for example, derivative observables) provided the $\tau$-ordering
and non-overlapping properties are respected. For well separated
observables no distinction between a quantum and classical $\tau$-ordered
correlation function would be needed. In particular, this holds also
for ``smoothened'' observables $A_i$ that involve (weighted) averages
over $\varphi(\tau)$ in a range ${\cal R}_i$ around $\tau_i$. Decreasing the
distance between $\tau_2$ and $\tau_1$, the new features of the quantum
product $A_1(\tau_2)\circ A_1(\tau_1)$ show up only once the distance
becomes small enough so that the two ranges ${\cal R}_1$ and ${\cal R}_2$
start to
overlap. In an extreme form the difference between quantum and classical
correlations becomes apparent for derivative observables at the same
location. Quite generally, the difference between the quantum and
classical product is seen most easily on the level of the
associated operators
\begin{eqnarray}\label{6.10}
A_1\circ A_2&\to& \hat A_1\hat A_2\nonumber\\
A_1\cdot A_2&\to& T(\hat A_1\hat A_2)\end{eqnarray}
Here $T$ denotes the operation of $\tau$-ordering. The $\tau$-ordered
operator product is commutative $T(\hat A_1\hat A_2)=T(\hat A_2\hat A_1)$
and associative $T(T(\hat A_1\hat A_2)\hat A_3)=
T(\hat A_1 T(\hat A_2\hat A_3))\equiv T(\hat A_1\hat A_2\hat A_3)$.
As we have seen in the discussion of the derivative observables,
it lacks, however, the general property of robustness with respect
to the precise definition of the observables. This contrasts with the
non-commutative product $\hat A_1\hat A_2$.
This discussion opens an interesting perspective: The difference
between classical and quantum statistics seems to be  a
question of the appropriate definition of the correlation function.
Simple arguments of robustness favor the choice of the quantum correlation!
This remark remains valid if we consider averaged or smoothened observables instead
of ``pointlike'' observables \cite{CWQ}.
In a sense, the successful description of nature by quantum-mechanical
operators and their products
gives an ``experimental indication'' that quantum correlations should be
used!

\section{Incomplete classical statistics, irrelevant and inaccessible
information}

Our discussion of incomplete classical statistics may perhaps have led to
the impression that the quantum mechanical properties are somehow related
to the missing information. This is by no means the case! In fact, our
investigation of the consequences of incomplete information about
the probability distribution was useful in order to focus the
attention on the question which information is really necessary to
compute the expectation values of local observables. We can now turn
back to standard ``complete'' classical statistics where the full
probability distribution $p[\varphi(\tau)]$ is assumed to be known.
We concentrate here on a general class of probability distributions
which can be factorized in the form $p=p_>p_0p_<$ according
to (\ref{1.1}) -- it may be called ``factorizable'' or
``$F$-statistics''. For example, all systems which have only
local and next-neighbor interactions are of this form. Within
$F$-statistics the states remain defined according to (\ref{2.7}).

We emphasize that any additional information contained in
$p[\varphi]$ which goes beyond the local distribution $p_0[\varphi]$ and the
states $|\psi\}$ and $\{\psi|$ does not change a iota in
the expectation values of local observables and their correlations!
The additional information is simply {\em irrelevant} for the
computation of local expectation values. A given probability
distribution specifies $p_<$ and $p_>$ uniquely. This determines
$|\psi\}$ and $\{\psi|$ and we can then
continue with the preceding discussion in order to calculate the
expectation values of local observables. The precise form of
$p_<$ and $p_>$ which has led to the given states plays no role
in this computation.

Since all information contained in $p_<$ and $p_>$ beyond the states
$|\psi\}$ and $\{\psi|$ is irrelevant for local expectation values,
it is also {\em inaccessible} by any local measurements. In fact,
even the most precise measurements of expectation values and
correlation functions for arbitrarily many local observables
could at best lead to a reconstruction of the states $|\psi\}$ and $\{
\psi|$. This sheds new light on the notion of ``incompleteness''
of the statistical information discussed in this note. In fact,
within $F$-statistics the ``incomplete'' information contained in
the states $|\psi\}$ and $\{\psi|$ constitutes the most complete
information that can possibly be gathered by local measurements! Since
any real measurement is local in time and space all assumptions
about information beyond the states concern irrelevant and
inaccessible information and cannot be verified by observation!

\bigskip
\section{Conclusions and discussion}

\medskip
Within a simple example of classical statistics for coupled unharmonic
oscillators on a chain we have formulated a description in terms of states and
operators in analogy to quantum mechanics. The state vectors and the operators
can be expressed in terms of classical functional integrals. Expectation values of
classical observables can be evaluated as ``quantum mechanical'' expectation values
of appropriate operators in appropriate states. Typical quantum
mechanical results like the relations between the expectation values in stationary
states or the uncertainty relation can be taken over to the classical system \cite{CWQ}.
The simple fact that quantum-mechanical information can be used in practice
to establish properties of expectation values in a standard classical statistical system
demonstrates in a simple way that quantum-mechanical features are indeed genuine properties
of classical statistical systems.
Our procedure inverts the construction of the Euclidean path integral for a
quantum mechanical system in the ground state or thermal state
\cite{Feynman} \cite{1a}, with a generalization to a wider class of states.

The introduction of ``quantum mechanical'' operators associated to every local classical
observable allows us to define equivalence classes of observables which cannot be
distinguished by any measurement of their expectation values. We argue that the definition
of the correlation function should be consistent with this equivalence structure. We
require that indistinguishable observables must lead to the same correlation function.
This leads to the introduction of a quantum correlation within the classical statistical
setting. We point out that the quantum correlation constitutes a more robust definition
of the correlation function with respect to the precise details of the definition of
observables, both for classical and quantum statistical systems.
The basic conceptual distinction between quantum statistics and classical statistics
disappears in this respect.
The similarity can be extended to the emergence of typical characteristics of quantum
statistics like the superposition of states and interference for classical
statistical systems \cite{CWQ}.
This raises the question \cite{EPR} if it could be possible to understand
the mysteries of the basics of
quantum mechanics within a formulation of a classical statistical
problem with infinitely many degrees of freedom.

\end{document}